# FACTORS ENHANCING E-GOVERNMENT SERVICE GAPS IN A DEVELOPING COUNTRY CONTEXT

Gilbert Mahlangu, Cape Peninsula University of Technology and Midlands State University, mahlago97@gmail.com

Ephias Ruhode, Cape Peninsula University of Technology, RuhodeE@cput.ac.za

**Abstract**: Globally, the discourse of e-government has gathered momentum in public service delivery. No country has been left untouched in the implementation of e-government. Several government departments and agencies are now using information and communication technology (ICTs) to deliver government services and information to citizens, other government departments, and businesses. However, most of the government departments have not provided all of their services electronically or at least the most important ones. Thus, this creates a phenomenon of e-government service gaps. The objective of this study was to investigate the contextual factors enhancing e-government service gaps in a developing country. To achieve this aim, the TOE framework was employed together with a qualitative case study to guide data collection and analysis. The data was collected through semi-structured interviews from government employees who are involved in the implementation of e-government services in Zimbabwe as well as from citizens and businesses. Eleven (11) factors were identified and grouped under the TOE framework. This research contributes significantly to the implementation and utilisation of e-government services in Zimbabwe. The study also contributes to providing a strong theoretical understanding of the factors that enhance e-government service gaps explored in the research model.

## 1. INTRODUCTION

E-government is the praxis of transforming government services from traditional to electronic means using modern information communication technologies (ICTs) to provide easy access to government services for all users such as citizens, businesses, and government agencies (Hanum et al., 2020). Globally, the discourse of e-government has gathered momentum in the public service delivery (Alabdallat, 2020; Almutairi et al., 2020; Jacob et al., 2019; Lee & Porumbescu, 2019; Lindgren et al., 2019; Mukamurenzi, 2019; Sánchez-Torres & Miles, 2017; Twizeyimana & Andersson, 2019). Accordingly, Alabdallat (2020) revealed that no country has been left untouched in the implementation of e-government. Several government departments and agencies are now using ICTs to deliver government services and information to citizens, other government departments, and businesses. However, Alabdallat (2020, p. 5) noted that "most of the government departments have not provided all of their services electronically or at least the most important ones. This issue seems to be confined to the developing countries, especially among countries with very low incomes". Thus, this creates a phenomenon of e-government service gaps.

There is a range of definitions of this term, but in this study, e-government service gaps refer to the extent to which e-government services are not fulfilled to the intended users (government employees, businesses and citizens) of the e-government system (Herdiyanti et al., 2018) either because the system is constrained to deliver the required services or some of the expected services are not being provided. Thus, e-government service gaps represent two major fine points: the constraints on the system to deliver e-services and the service deficiencies from the government.

E-government service gaps in developing countries have not yet been bridged (Sterrenberg & Keating, 2016). Despite the significant amounts of public investment committed to enhancing e-government over the past two decades, the use of this service is still limited (Pérez-Morote et al., 2020). Citizens are still required to visit respective government departments and agencies to get basic information, complete and submit forms (Madariaga et al., 2019) or get other services that possibly can be offered electronically (Alraja, 2016; Khamis & Weide, 2017). Predominantly, public service delivery in developing countries is still characterised by inefficient, rigid and manual systems (Singh & Travica, 2018; Yang, 2017).

This study aimed to explore the opinions of government employees, businesses and citizens to understand the contextual factors enhancing e-government service gaps in a developing country from multiple perspectives. To accomplish the objective of the study, the primary research question was formulated as follows:
*What are the contextual factors enhancing e-government service gaps in a developing country resulting in low usage?*

## 2. LITERATURE REVIEW
### 2.1. The concept of e-government
Ideally, e-government is expected to decrease travelling costs, reduce waiting time for the service, reduce operational time, decrease corruption and cost in service delivery, increase transactional capabilities and convenience and improve accessibility (Alabdallat, 2020; Dewa & Zlotnikova, 2014; Kalu & Masri, 2019; Nabafu & Maiga, 2012). The emphasis on e-government implementation is put on how to transform both internal and external relationships of governments to reduce complex and over-stretched bureaucratic system (Alassim & Alfayad, 2017; Almutairi et al., 2020; Janowski, 2015; Mees et al., 2019). The emergence of e-government was a response to make government departments and agencies more efficient and open in public service delivery by utilising ICTs to provide services electronically (Sharma, Bao & Peng, 2014). Thus, the resulting transformation makes the government more efficient and transparent in delivering public services. For these reasons, the concept of e-government is treasured for being evolutionary, transformational, efficient, and transparent.

### 2.2. E-government service delivery models
Accordingly, the transformation drive in public service is facilitated by the following e-government delivery models: Government-to-Government (G2G); Government-to-Employees (G2E); Government-to-Business (G2B); and Government-to-Citizens (G2C) (Ahmad et al., 2019; Ramdan et al., 2014; Voutinioti, 2014). On the first hand, G2G represents the backbone platform for e-government adoption, implementation and utilisation in the entire country (Voutinioti, 2014). On the other hand, G2E represents an internal relationship between the government and its employees (Ramdan et al., 2014). G2B service delivery model denotes an online platform that enables government and business organisations to do business electronically (Ahmad et al., 2019). Lastly, G2C ensures that the citizens interact and transact with the government far and wide (Ramdan et al., 2014).

### 2.3. E-government in developing countries: The African context
Studies have confirmed that most of the government services in developing countries particularly in Africa are still unavailable online (Humphrey et al., 2016; Rabaa et al., 2018; Sarrayrih & Sriram, 2015; Singh & Travica, 2018; Twizeyimana et al., 2018). African region lags in e-government development compared to the rest of the world. Basic e-government services are still not easy to find in African countries; only limited services are offered online (Agboh, 2017; Mukamurenzi, 2019; Owusu-Ansah, 2014; Twizeyimana et al., 2018; Verkijika & De Wet, 2018). Most government services are still provided manually. Probably, this is the reason why developing countries have the lowest e-government service development intensity. Thus, this suggests that there are e-government service gaps that have resulted in low usage of e-government and limited capabilities of the government employees to provide efficient services (Tirastittam et al., 2018).

## 2.4. E-government services in Zimbabwe: The research context

Like many other developing countries, Zimbabwe is implementing e-government projects as part of the public sector reforms to improve service delivery among government departments, agencies, businesses and citizens (Munyoka, 2019; Nhema, 2016). The major target of the public sector reforms in the context of e-government is to exterminate institutional constrictions related to the conventional methods of public service delivery. Even though the country seems to be committed to the implementation of e-government projects, still, the ability to provide comprehensive e-services is not attainable. The country compares relatively low with other countries in the world, the African continent as well as in the Southern African Development Committee (SADC) region (Munyoka, 2019). In fact, by 2018, the country was ranked 146 out of 193 in EGDI and last in SADC while the Online Service Index (OSI) stood at the mean value of 0.3246 (Dias, 2020). This ranking reflects a significant need in providing comprehensive e-government services.

## 2.5. Factors affecting e-government adoption, implementation and usage

E-government projects in developing counties face a variety of challenges during their implementation and utilisation but the severity of these issues varies from context to context (Mustafa et al., 2020). Generally, these challenges can be categorised as follows:

### 2.5.1. Infrastructure

Many studies have concluded that developing countries do not have adequate infrastructure to successfully deploy e-government projects (Baheer et al., 2020; Hanum et al., 2020; Heeks, 2003; Kanaan et al., 2019). Challenges such as low penetration of fixed-line telecommunications; inadequate electricity supply (Richardson, 2011) and low teledensity (Sareen et al., 2013) make it difficult to deploy e-government countrywide. This condition has resulted in e-government service gaps. The gaps are created in multifold: first and foremost, lack of infrastructure hinders the delivery of e-government services by acting as an obstruction for government departments and agencies to provide e-services; secondly, lack of infrastructure obstructs the demand for e-government services by impeding citizens to access e-government services; and lastly, unreliable infrastructure can degrade the performance of e-government systems; thereby, making it difficult to for users to obtain higher-level e-government services.

### 2.5.2. Interoperability

In the context of e-government, interoperable depicts the ability of independent systems and devices to communicate with each other and share data (Apleni & Smuts, 2020; Sulehat & Taib, 2016). Most of the e-government systems deployed in developing counties operate in 'silos'; the e-government landscape is fragmented within and across ministries, departments and agencies (Apleni & Smuts, 2020; Baheer et al., 2020; Mohlameane & Ruxwana, 2020; Nakakawa & Namagembe, 2019; Sulehat & Taib, 2016). This situation has made the realisation of e-government benefits merely a delusion. In consequence, the lack of interoperability results in the loss of entirely reaping the prospective benefits of e-government such as more efficiency; enhanced services to better serve citizens; and better accessibility of public services.

### 2.5.3. Digital divide

The digital divide is a dynamic and complex problem that is creating service gaps in developing countries particularly in the utilisation of e-government services. It is the gap between people who have access to the internet and those who do not (Alabdallat, 2020). The digital divide reflects the lack of and/or limited access to electronic services by citizens. It is regarded as a significant barrier to the implementation and utilisation of e-government since many communities and citizens do not have access to the internet and computing devices (Alabdallat, 2020; Chipeta, 2018; Idoughi &

Abdelhakim, 2018; Twizeyimana & Andersson, 2019b). This restricts the adoption and utilisation of e-government to those who have access to the technology and the requisite skills to use e-services. Therefore, those who do not have access to ICTs and necessary ICT skills cannot access e-services; and thus fail to benefit from e-government projects implemented in their service constituencies (Haider et al., 2015; Twizeyimana & Andersson, 2019; Verkijika & De Wet, 2018).

### 2.5.4. Human factor
The human factor is critical in the success of e-government. According to Farzianpour et al. (2015), once the infrastructure has been established, there is a need for ICT skills to enhance the effective implementation and utilisation of online services. Nevertheless, a range of studies has reported that the lack of ICT skills is the dominant human aspects under the barriers to e-government initiatives (Aneke, 2019; Khan & Ahmad, 2015; Owusu-Ansah, 2014). For instance, Owusu-Ansah (2014) reported that e-government has failed in developing countries due to inadequate ICT skills among government employees and citizens. Apart from lack of skills, various studies on critical success factors in the implementation of e-government in developing countries have reported the lack of expertise by government employees to develop, operate and maintain e-government systems (Aneke, 2019; Khadaroo et al., 2013; Layne & Lee, 2001).

### 2.5.5. Policy factor
According to Dias (2020), a policy is a premeditated plan of action aimed at guiding decisions and accomplishing judicious outcomes. The issue of policy as well as forms part of the factors that affect the implementation of e-government (Apleni & Smuts, 2020). This is because the deployment and use of e-government systems call for a variety of policies to regulate electronic activities. However, Islam (2013) noted with great concern that in developing countries there is a lack of clearly defined policy for e-government implementation. Very few countries (Singapore and Malaysia) have stand-alone policies for implementing e-government (Alabdallat, 2020; Apleni & Smuts, 2020; Bwalya, 2009; Dias, 2020; Nurdin et al., 2014; Zaied et al., 2017); the implementation of e-government is either driven by national ICT policies or it is the sole responsibility of the ministries, departments and agencies (MDAs). This factor demonstrates a major policy gap in the implementation of e-government projects in developing countries.

### 2.5.6. Funding
Funding is the priority factor for successful e-government adoption because "any e-government initiatives require funding to initiate and maintain e-government projects" (Apleni & Smuts, 2020, p. 19). However, most developing countries are struggling to fund their e-government initiatives except for few countries. They lack financial support in the implementation of e-government projects resulting in a funding dilemma even if governments have plans for the implementation of e-government (Fasheyitan, 2019; Khadaroo et al., 2013; Ziba & Kang, 2020). As a result, most e-government projects particularly in African countries are donor-funded. The reliance on donor support for e-government implementation often results in untenable funding in the event donor support is terminated; thus, impeding progress in the implementation of e-government (Khadaroo et al., 2013).

## 3. THEORETICAL FRAMEWORK
This study used the Technological- Organisational- Environmental (TOE) model to study the factors enhancing e-government service gaps in a developing country context. According to Defitri et al. (2020), "the TOE framework originates from the theory of adoption of new technologies, making [it] widely adopted in various studies as compared to other models" (p. 40). This framework provides key benefits for understanding the factors that exist in the context of technology, organisation, and environment in influencing the process of adopting technological innovation. Technological factors refer to the existing and new technologies, for example, telecommunications, electricity, computers, and bandwidth necessary for the implementation and usage of e-government services while the organisational factors describe the attributes of an organisation such as top management support, e-government funding, skills and competences. The environmental context focuses on the external factors that drive the adoption and implementation of new technology policy and regulations. Furthermore, the TOE was adopted as the underpinning theory in this study because it was considered to unveil a wide range of factors associated with technology, organisation and the environment that may shape the implementation and usage of e-government services in a developing country. Also, as a generic theory, the TOE enabled researchers to choose the factors of each facet according to the characteristics of technological innovation (e-government) adopted. Thus, the analytical framework in Figure 1 forms the basis for investigating the factors enhancing e-government service gaps.

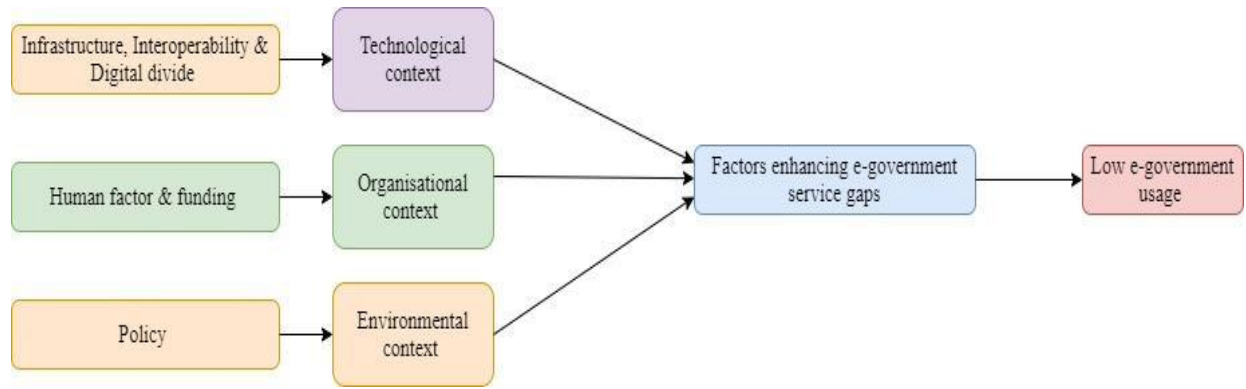

**Figure 1. The analytical framework for the study**

## 4. METHODOLOGY

A multiple-case study was adopted as the research strategy of the study in which the e-Taxation system was the primary case study with embedded three (3) units of analysis (government employees, business, and citizens). These cases were used as the primary sources of empirical data because it was assumed that they were adequate to provide sufficient data about the study (Yin, 2009). The selection of cases began by recognising the key units that could benefit directly from the deployment of e-government projects in Zimbabwe. The study used a multi-stage sampling procedure by employing stratified, purposive and snowball sampling techniques for the following reasons:
- Stratified sampling ensured that each subunit of analysis was represented in the sample.
- Purposive sampling also enabled researchers to select participants that showed the best ability to address the research question and to meet the research objectives of the study.
- Snowball sampling was used because the citizens who are users of the e-Taxation system were very difficult to find.

A total of 30 semi-structured interviews were conducted with government employees, managers of various business organisations and citizens who have direct contact with the e-Taxation system. The selected size was informed by Weller et al. (2018) who reported that 20 to 30 in-depth interviews are adequate to determine the saturation point of qualitative data. This study employed a template analysis technique proposed by King (2004) to code qualitative data from interviews transcripts. Template analysis is a technique within the thematic analysis framework that enabled researchers to develop a catalogue of codes to form an analytic template representing *priori* themes identified from the literature and presented in the analytical framework (Brooks & King, 2014).

The study focused on multiple stakeholders (users) to uncover multiple realities on the implementation and use of e-government in a developing country. From the onset, the researchers assumed that using a single stakeholder perspective would not reveal numerous factors enabling e-government service gaps. Accordingly, by focusing on multiple perspectives, the study diverted from previous studies which have investigated e-government in isolation by focusing their effort on a single perspective. Mostly, e-government studies have been centred on citizens' perspectives despite the importance of other user groups. Also, e-government projects are designed and implemented to satisfy three major users: citizens, government employees and businesses (Haider et al., 2015; Ibrahim et al., 2016). Besides, these users have different opinions and values about e-government, too, which could result in different perspectives. Therefore, e-government evaluation should recognise divergent views of users.

## 5. FINDINGS OF THE STUDY

This section reports the findings from three (3) units of analysis on factors enhancing e-government service gaps in a developing country like Zimbabwe. The findings were presented using the thematic configuration (structure) of the template, explaining the meaning of themes, and illustrating with direct quotes from the data (Ryan & Bernard, 2003). The factors were coded and classified using template analysis based on three theoretical codes: technology, organisation, and environment. The factors are presented below.

| Codes | Themes | Supporting excerpts |
|---|---|---|
| Technology | Lack of government-owned infrastructure | *"In the case of Zimbabwe, the private sector owns approximately 75% of the total infrastructure deployed across the country"* [Bus 1]. |
| | Lack of compatibility of computing devices | *"Some of our gadgets hinder the electronic exchange or transfer of information as they are not compatible with the technology used in e-government projects"* [Bus 2]. |
| | Lack of systems integration | *"Most government departments use legacy systems that cannot easily integrate with other systems"* [CIT 8]. |
| | Lack of electricity infrastructure | *"The country lacks adequate electricity supply; electricity load shedding is very high in Zimbabwe, resulting in citizens failing to access internet services"* [Bus 6]. |
| | Lack of access | *"Several communities in rural areas are still underserved as far as internet connectivity is concerned; hence, they do not have access to the internet"* [CIT 4]. |
| Organisation | Lack of e-government funding | *"There is a lack of funding for e-government projects as developing countries like Zimbabwe have many competing priorities to finance"* [Govt employee 5]. |
| | Budget disparity | *"There is a lot of budgetary politics in the government of Zimbabwe; some government departments do not get sufficient budget"* [Govt employee 7]. |
| | Lack of the desire to support and coordinate e-government | *"The top management is not 'aggressively' lobbying for adequate resources to facilitate the deployment of robust infrastructure"* [Govt employee 3]. *"Each Ministry or government department is concentrating on developing its system and deploying individual infrastructure without considering the need to coordinate such activities with other departments or even the private sector"* [Govt employee 7]. |
| | Design-reality gap | *"In Zimbabwe, I think e-government service gaps exist because of skill gaps among e-government designers. Most of the e-government designers have limited knowledge and experience"* [CIT 9]. |
| Environment | Policy inconsistency | *"There is a tendency to jump from policy to policy such that policy pronunciation becomes inconsistency with goals and aspirations of e-government"* [Govt employee 12]. |
| | Lack of user-involvement | *"There is an assumption that the designers of e-government systems know all the needs and expectations of the users in advance. The end users are not consulted during the design phase"* [CIT 4]. |

**Table 1. Codes, themes and supporting excerpts**

## 6. DISCUSSION OF FINDINGS

### 6.1. Technology
#### 6.1.1. Lack of government-owned infrastructure
It is worth mentioning that the single most striking observation to emerge from the empirical data compared to previous studies was the lack of government-owned infrastructure. Still, in cases where the government owns the infrastructure, it is dilapidated and not able to match the needs of the modern-day e-government service provision. Drawing on the analysis of qualitative findings the government of Zimbabwe has not invested significantly in ICT infrastructure and largely relies on infrastructure from private players. This was echoed across cases:

Most of the infrastructure in Zimbabwe is not timely updated to suit the current needs of the end-users" [Bus 3].

"In the case of Zimbabwe, the private sector owns approximately 75% of the total infrastructure deployed across the country whilst the government does not have adequate resources to set up the infrastructure for e-government" [Bus 1].

"In terms of the ownership of ICT infrastructure, the private sector has more ownership compared to most government departments" [Govt employee 8].

This implies that the participants were concerned with the lack of government-owned infrastructure because relying largely on infrastructure from the private players becomes expensive and unaffordable for the government to sustainably run e-government schemes.

### 6.1.2. Lack of electricity infrastructure

The users of e-government are failing to access e-government due to inadequate electricity supply. The respondents were concerned with the lack of reliable electricity supply which could inhibit the continuous delivery of e-government services. The following excerpts form part of the narratives that were used to develop the theme of lack of electricity infrastructure:

"Power supply availability is a challenge in Zimbabwe. Most places do not have a reliable power supply infrastructure for offices; this makes it difficult to provide e-government services" [Govt employee 13].

"The country lacks adequate electricity supply; electricity load shedding is very high in Zimbabwe, resulting in citizens failing to access internet services" [Bus 6].

"Electricity load shedding is very high in Zimbabwe. Even in town, there are continuous electricity/power cuts; this makes e-government services not available on time" [CIT 7].

This finding implies that without a stable electricity supply, e-government systems cannot run smoothly and user satisfaction is likely to be minified since the system might tend to be ordinarily offline. This claim seems to be plausible in the context of a developing country where electricity supply has remained a challenge for decades (Richardson, 2011; Ud Din et al., 2017). Thus, this gives credence to the fact that electricity is fundamental in the implementation and utilisation of e-government.

### 6.1.3. Lack of systems integration

Integration refers to the extent to which e-government systems can share information to enable citizens to access services from various departments and agencies using a single access point (Dias, 2020). In the literature, integration is identified as a key challenge for enabling the fully functional and higher maturity level of e-government in developing countries (Bayona & Morales, 2017; Owusu-Ansah, 2014). The participants across cases mentioned that e-government systems that have been deployed in government departments are not interoperable:

"System integration is lacking in e-government systems as most systems exist in 'silos'. There is silo mentality in systems development, every MDA is concerned with the services it offers; hence, ignoring the need for system integration" [Govt employee 8].

"Every government department has its system which is not integrated with other departments; thereby, creating e-government service gaps" [Govt employee 13].

"E-government systems in Zimbabwe fail partly because they have been designed in such a way that they do not communicate with other systems deployed across ministries. Most government departments use legacy systems that cannot easily integrate with other systems" [CIT 8].

This implies that systems integration is one of the issues within the e-government domain that need to be managed by any government intending to derive added value from e-government initiatives (Nakakawa & Namagembe, 2019; Sulehat & Taib, 2016). Thus, without systems integration, MDAs that support each other will find it difficult to share critical data and information.

### 6.1.4. Lack of access
Research has shown that lack of access to the internet and technology is a cause for concern in the utilisation of e-government services in developing countries (Abu-Shanab & Khasawneh, 2014; Ohemeng & Ofosu-Adarkwa, 2014; Regmi, 2017). The majority of the population, particularly in developing countries, still has limited or no access to e-government services even though these countries have moved a great stride in e-government adoption. Similarly, the findings across cases revealed that the majority of citizens in Zimbabwe do not have access to the internet, e-government services and computing devices due to poverty, low levels of income and inadequate network coverage.
"Lack of access to the internet is also another factor; the charges for internet connectivity and data are beyond the reach of many ordinary citizens" [Bus 5].

"Several communities in rural areas are still underserved as far as internet connectivity is concerned; hence, they do not have access to the internet" [CIT 4].

"Most people in rural areas are poor and lag in terms of digitisation. Some of them could not afford a Smartphone, not even talking about the data bundles" [CIT 8].

This implies that those who do not have access to computers, network coverage or the internet are unable to benefit from e-government services. However, it should be noted that there is an interlocked relationship between infrastructure and access to e-government services. Lack of infrastructure results in lack of access because it deprives citizen to benefit from the implementation of e-government; hence, creating a phenomenon of e-government service gaps.

### 6.1.5 Lack of compatibility of computing devices
Most research on compatibility has focused on measuring the degree to which technology is consistent with the present values, demands and previous experiences of the prospective users (Abu-Shanab & Khasawneh, 2014; Ahmad & Campbell, 2015; Dhillon & Laxmi, 2015; Kumar et al., 2007; Layne & Lee, 2001; Muhammad, 2013; Zautashvili, 2018). However, this study provides a new understanding of the compatibility dimension in which it refers to the ability of different computing devices to access e-government systems. This is because users of e-government use different gadgets; so there is a need for assurance that they may be able to have access to the system despite using different computing devices.
"Some of our gadgets hinder the electronic exchange or transfer of information as they are not compatible with the technology used in e-government projects" [Bus 2].

Therefore, the compatibility of computing devices should be significantly considered in the development and deployment of e-government systems to provide user-centric services. Thus, e-government systems should be compatible with the computing devices of the users.

### 6.2. Organisation
### 6.2.1. Lack of e-government funding
It is a widely held view that adequate funding is the factor that promotes the success of e-government implementation (Alabdallat, 2020; Alanezi et al., 2012; Khadaroo et al., 2013). The availability of sufficient funding is a significant factor towards the successful implementation of e-government because there is a strong correlation between funding and ICT infrastructure development, addressing the digital divide and human capacity development (Khadaroo et al., 2013; Ziba & Kang, 2020). The empirical data reveals that Zimbabwe is facing difficulties in financing e-government projects due to competing priorities to finance; for instance, funding the everyday needs of the citizens in which the majority survive on government hand-outs.

"There is a lack of funding for e-government projects as developing countries like Zimbabwe have many competing priorities to finance" [Govt employee 5].

"Many developing countries fall into a dilemma of funding e-government programs due to lack of financial support even the government entity has a plan for providing e-government services" [Govt employee 2].

"There is a lack of funding in Zimbabwe to acquire the necessary infrastructure required for the full implementation of the e-government; the central government seems to have many other priorities competing for the resources" [Govt employee 7].

The findings of the current study are consistent with those of Alabdallat (2020) who reported that there is a lack of funding for e-government projects in developing countries due to many competing priorities to finance from constrained budgets.

### 6.2.2. Budget disparity

This study defines budget disparity as a discrepancy in the allocation of budget among the government departments; as a result, some departments are receiving lesser budgets than others. The case study findings point that successful implementation of e-government requires sufficient budget across government departments; otherwise, other departments will lag in the implementation of e-government projects; hence, resulting in e-government service gaps.

"There is a lot of budgetary disparity in the government of Zimbabwe; some government departments do not get sufficient budget because they are neither preferred nor favoured" [Govt employee 7].

"I have noted that the abilities of government departments to place services online and to use technology for automating processes are hampered by budget disparity" [Govt employee 5]

"At the same time, resources differ from ministry to ministry because some ministries receive better budgets than others; therefore, we cannot be at the same level in the implementation of e-government" [Govt employee 3].

The finding implies that government departments will always be at different levels of e-government maturity. This study is consistent with the literature that discusses the impact of budget on e-government implementation in general (Dhonju & Shakya, 2019), but not necessarily specific to the budget disparity.

### 6.2.3. Lack of the desire to support and coordinate e-government

In this study, the participants indicated that there is a lack of top management support and coordination in the deployment of e-government projects. Each ministry or government department concentrates on developing its system and deploying individual infrastructure without considering the need to coordinate such activities with other departments or even the private sector. The lack of coordination in systems development is likely to lead to MDAs competing in putting up ICT infrastructure while lack of coordination can restrain the establishment of appropriate e-government development networks.

"Lack of top management support ... has resulted in unavailability of the requisite infrastructure; the top management is not 'aggressively' lobbying for adequate resources to facilitate the deployment of robust infrastructure" [Govt employee 3].

"Lack of coordination in systems development lead to MDAs competing in putting up ICT infrastructure; hence, there is a lot of fragmented (silo) efforts in the deployment of e-government in Zimbabwe" [Govt employee 1].

This implies that the implementation of e-government in Zimbabwe is among factors hindered by the lack of the desire by the top management to support and coordinate the design and deployment of e-services.

### 6.2.4. Design-reality gap

The successful implementation of e-government projects demands the fusion of IT human capacities for designing, installing, maintaining and utilising e-government systems. The case study findings revealed that one of the challenges in e-government design is the lack of ICT skills and experience in government departments due to poor remunerations for IT personnel. The majority of skilled and experienced IT personnel are either employed in the private sector or in other countries. Most of the e-government designers have limited knowledge and experience; as a result, developing systems that are not 'perfect' and which do not meet the needs and expectations of the users. Businesses and citizens shared similar views regarding the design-reality gap.

"In Zimbabwe, I think e-government service gaps exist because of skill gaps among e-government designers" [CIT 9].

"Skills gap will always be there by virtue of our maturity in terms of e-government and poor remuneration" Development of such complex systems is mostly outsourced from consultants who have expertise in developing similar systems in other countries" [Bus 4].

This is why in developing countries we find private companies with better e-services than governments" [CIT 6].

"The challenge is that government departments do not offer motivating remuneration and working conditions that are conducive to drive the implementation of e-government to success. Most skilled employees end up leaving for greener pastures in the private sector and abroad" [CIT 3].

It appears that incompetent employees are appointed to develop and maintain e-government systems in Zimbabwe. As a consequence, e-government projects are outsourced from developed countries which according to Heeks (2003) fuels the design-reality gap if the project is adopted in its entirety by a developing country. So, e-government cannot be successfully deployed if government employees do not have adequate ICT skills and experience.

### 6.3. Environment
### 6.3.1. Policy inconsistency

Another drawback in the successful implementation of e-government in a developing context is policy inconsistency. This factor demonstrates a major policy gap in the implementation of e-government projects in developing countries. During the interviews, participants from the government stratum noted that the government of Zimbabwe always grapples with policy inconsistency.

"One other drawback is the policy inconsistency that the government of Zimbabwe grapples with; so nobody takes the government seriously when it outlines policies" [Govt employee 6].

"There is only too much rhetoric and very little traction on the factors obtaining on the ground in the implementation of e-government projects" [Govt employee 13].

"There is a tendency to jump from policy to policy such that policy pronunciation becomes inconsistency with goals and aspirations of e-government" [Govt employee 12].

"Midway into the implementation of one policy, the government usually shift goals and launch another policy to the puzzlement of e-government designers; leading to 'half-done and later deserted projects" [Govt employee 11].

Although the country strives to digitise the public sector by coming up with e-government implementation policies, this aspiration has remained a mere declaration of the intent; in that respect, the obligatory vigour to drive the implementation of e-government in Zimbabwe is still missing at all levels. There is too much rhetoric and very little traction in the implementation of e-government projects. Besides, it is noted that in most cases, governments take more than a decade talking about implementing e-government flagship projects which do not take off as expected.

### 6.3.2. Lack of citizen involvement

The scoping of any information systems project requires the involvement of users. The main consideration of citizen involvement is to incorporate their opinions in the design of e-government projects. The findings from the citizen stratum, however, revealed that the e-government design phase is not engaging with the citizens. As a consequence, designers are likely to develop e-government schemes that do not result in user acceptance and satisfaction.

"There is an assumption that the designers of e-government systems know all the needs and expectations of the users in advance. The end users are not consulted during the design phase; hence, at times they resist the adoption of e-government schemes" [CIT 4].

"The other challenge with most e-government schemes is that they are developed at the national level and lack engagement of the users who are supposed to benefit from such schemes. The users are not engaged; hence the failure of the users to embrace these schemes" [CIT 2].

"The e-government design phase is not engaging with the citizens. The opinions of citizens in the design of e-government are not incorporated; that is why it is not easily accepted" [CIT 5]. "

This implies that e-government projects fail largely due to the non-factoring of the individual requirements and needs during the design and implementation of e-government projects (Ahmad et al., 2012). The present findings seem to be consistent with other research which found out that user involvement is not prioritised in current e-government development projects (Abu-Shanab & Khasawneh, 2014). Similarly, Verkijika (2018) declared that e-government projects fail to provide comprehensive services because of a lack of engagement with the users to capture their needs and wants in the design phase.

## 7. CONCLUSIONS AND RECOMMENDATIONS

The study explored the opinions of government employees, businesses and citizens to understand the factors enhancing e-government service gaps in a developing country context from multiple perspectives underpinned by the TOE framework. The study identified 11 factors that deeply enhance e-government service gaps in a developing country. Since the study was underpinned in the TOE the factors that emerged from the findings were grouped under the TOE elements. These factors can be regarded as the baseline elements in the implementation and utilisation of e-government services because they act as barriers for governments to successfully implement e-government and prevent users to engage with e-government services. Furthermore, the factors hold both negative and positive outcomes; mostly because apiece, they have generative mechanisms to make a divergence of either enhancing or constricting e-government service gaps. Therefore, this study concluded that until the factors enhancing e-government service gaps are converted into enablers (enabling factors) for providing comprehensive services, e-government service gaps will continue to exist in developing countries.

While factors enhancing e-government service gaps may be contextual, a systematic literature review shows that most e-government challenges in the implementation and usage of e-government services fall in the wider factors identified in this study. The findings of this study show that lack of government-owned infrastructure, systems integration, e-government funding, the desire to support and coordinate e-government; design-reality gap and policy inconsistency are the critical factors enhancing e-government service gaps in Zimbabwe.

However, it should be noted that while the aforesaid factors were derived from the three units of analysis others were derived solely from each unit or two units. For instance, the lack of government-owned infrastructure was extrapolated from all units while budget disparity; lack of the desire to support and coordinate e-government; and policy inconsistency were derived from the government stratum only. This could be attributed to the fact that these elements appear to be internal factors with a direct impact on government departments and their employees in their effort to drive the implementation of e-government. The lack of user involvement was deduced from the citizen stratum. In contrast, the design-reality gap was derived from business and citizen strata; thus, showing differences in the discernment of factors enhancing e-government service gaps among the strata.

The study provides the following recommendations:
- developing countries need to have a dedicated budget to fund the deployment of infrastructure and e-government projects;
- government departments should compete with the private sector in the ICT job market and offer better remuneration that keeps the best ICT employees working for their governments and drive e-government services;
- ensure that citizens are part of the design phase by involving users in the design of e-government systems since the designers of e-government alone cannot fully comprehend the needs and expectations of the users;
- there is a need to convert factors enhancing e-government service gaps to e-government enabling factors;
- priority should be raised in deploying government-owned infrastructure; e-government funding; and IT human capacity development

# 8. CONTRIBUTIONS TO PRACTICE AND THEORY

This research contributes significantly to the implementation and utilisation of e-government services in Zimbabwe, as well as contributing to knowledge on e-government implementation in developing countries, particularly in Africa. Furthermore, the research provides a framework to assist governments in developing countries to surmount the challenges of e-government service gaps. This study will also enlighten the implementers and funders of e-government projects on factors that obstruct the successful implementation and utilisation of e-government services in a developing context which is known for highly failed e-government projects.

The study contributes to providing a strong theoretical understanding of the factors that enhance e-government service gaps explored in the research model. From the extant literature, no studies have explicitly focused on investigating factors that enhance e-government service gaps in the context of a developing country. To fill this gap, the study adopted the TOE framework for better understanding these factors. This study contributes to theory by classifying e-government service gaps into three (3) categories: Technology, Organisation and Environment. Thus, the findings provide theoretical contributions to the body of knowledge concerning the factors that contribute to e-government service gaps in a developing country.

# 9. LIMITATIONS AND SUGGESTED AREAS FOR FURTHER RESEARCH

The factors enhancing e-government service gaps were proposed based on the literature review and explored using a single-embedded case study; hence, it is hard to conclude that the factors revealed in this study are conclusive. Furthermore, this study was focused on the urban population in Harare, Bulawayo and Gweru where the population has access to the internet and e-government experience in the use of e-government services. Therefore, by focusing on the urban population the study could not get the views of the non-users of e-government who are likely to experience more service gaps compared to the urban population.

The study suggests that new insights on factors enhancing e-government service gaps could emerge if the research is undertaken again in more case studies. Furthermore, the study suggests that future research on e-government service gaps should include the marginalised communities.

## ACKNOWLEDGEMENTS
Our special thanks are extended to organisations, businesses and citizens for their assistance with the collection of the empirical data. Their contributions are sincerely and greatly appreciated and were of vital significance to the success of this study.

### Authors' contributions
Both authors contributed equally towards the research and the writing of the article.

### Funding information
The authors declare that they did not receive grants from any funding agency in the public, commercial or not-for-profit organisations.

### Data availability statement
The data that support the findings of this study are available from the corresponding author, M.G., upon reasonable request.

# REFERENCES

Abbas, A., Faiz, A., Fatima, A., & Avdic, A. (2017). Reasons for the failure of government IT projects in Pakistan: A contemporary study. 14th International Conference on Services Systems and Services Management, ICSSSM 2017 - Proceedings. https://doi.org/10.1109/ICSSSM.2017.7996223

Abu-Shanab, E., & Khasawneh, R. (2014). E-government adoption: The challenge of digital divide based on jordanians' perceptions. Theoretical and Empirical Researches in Urban Management, 9(4), 5–19.

Abu-Shanab, E., Khasawneh, R., Li, Y., Shang, H., Herdiyanti, A., Adityaputri, A. N., Astuti, H. M., Song, W., Divide, D., Yarimoglu, E. K., June, G., Author, T., Reserved, A. R., Development, P., Thomson, L., Wimmer, M., Traunmüller, R., S, B. K., Featherman, M. S., … Annex, T. (2014). Developing Service Quality Using Gap Model-a Critical Study. IOSR Journal of Business and Management, 7(2), 92–100. https://doi.org/10.13189/aeb.2019.070204

Agboh, D. (2017). An Assessment of Ghana's global E-government UN ranking. Journal of Technology Research, 8,1–17.

Ahmad, K. M., & Campbell, J. (2015). Citizen perceptions of e-government in the Kurdistan Region of Iraq. Australasian Journal of Information Systems, 19, 1–29. https://doi.org/10.3127/ajis.v19i0.1201

Ahmad, K. M., Campbell, J., Pathak, R. D., Belwal, R., Singh, G., Naz, R., Smith, R. F. I., Al-Zoubi, K., Carter, L., Bélanger, F., Lim, A. L., Masrom, M., Din, S., Chen, Y. C., Dimitrova, D. V., Dodeen, W., Adolph, A., D., A., Bortier, S., … Munyoka, W. (2019). Satisfaction with e-participation: A model from the Citizen's perspective, expectations, and affective ties to the place. African Journal of Business Management, 7(1), 157–166. https://doi.org/10.1007/978-3-642-22878-0_36

Al-Azizi, L., H. Al-Badi, A., & Al-Zrafi, T. (2018). Exploring the Factors Influencing Employees' Willingness to Use Mobile Applications in Oman: Using UTAUT Model. Journal of E-Government Studies and Best Practices, 2018, 1–27. https://doi.org/10.5171/2018.553293

Al-Ghaith, W., Sanzogni, L., & Sandhu, K. (2010). Factors Influencing the Adoption and Usage of Online Services in Saudi Arabia. The Electronic Journal of Information Systems in Developing Countries, 40(1), 1–32. https://doi.org/10.1002/j.1681-4835.2010.tb00283.x

Alabdallat, W. I. M. (2020). Toward a mandatory public e-services in Jordan. Cogent Business and Management, 7(1). https://doi.org/10.1080/23311975.2020.1727620

Alanezi, M. A., Mahmood, A. K., & Basri, S. (2012). E-government service quality: A qualitative evaluation in the case of Saudi Arabia. Electronic Journal of Information Systems in Developing Countries, 54(1), 1–20. https://doi.org/10.1002/j.1681-4835.2012.tb00382.x

Alassim, M., & Alfayad, M. (2017). Understanding Factors Influencing E-Government Implementation in Saudi Arabia from an Organizational Perspective. 11(7), 894–899.

Albar, Mooduto, H. A., Dahlan, A. A., Yuhefizar, Erwadi, & Napitupulu, D. (2017). E-government service quality based on e-GovQual approach case study in West Sumatera province. International Journal on Advanced Science, Engineering and Information Technology, 7(6), 2337–2342. https://doi.org/10.18517/ijaseit.7.6.4226

Almutairi, F. L. F. H., Thurasamy, R., & Yeap, J. A. L. (2020). Historical Development of E-Government in the Middle East. International Journal of Recent Technology and Engineering, 8(5), 748–751. https://doi.org/10.35940/ijrte.e4912.108520

Alraja, M. N. (2016). the Effect of Social Influence and Facilitating Conditions on E-Government Acceptance From the Individual Employees' Perspective. Polish Journal of Management Studies, 14(2), 18–27. https://doi.org/10.17512/pjms.2016.14.2.02

Aneke, S. (2019). Challenges to e-government implementation in developing countries . Nigeria case Study. 3(2).

Apleni, A., & Smuts, H. (2020). An e-Government Implementation Framework: A Developing Country Case Study. In Lecture Notes in Computer Science (including subseries Lecture Notes in Artificial Intelligence and Lecture Notes in Bioinformatics): Vol. 12067 LNCS. Springer International Publishing. https://doi.org/10.1007/978-3-030-45002-1_2

Asogwa, B. E. (2011). The state of e-government readiness in Africa : A comparative web assessment of selected African countries. 2(October), 43–57.

Baheer, B. A., Lamas, D., & Sousa, S. (2020). A Systematic Literature Review on Existing Digital Government Architectures: State-of-the-Art, Challenges, and Prospects. Administrative Sciences, 10(2), 25. https://doi.org/10.3390/admsci10020025

Bayona, S., & Morales, V. (2017). E-government development models for municipalities. Journal of Computational Methods in Sciences and Engineering, 17(S1), S47–S59. https://doi.org/10.3233/JCM-160679

Bwalya, K. J. (2009). Factors Affecting Adoption of e-Government in Zambia. The Electronic Journal of Information Systems in Developing Countries, 38(1), 1–13. https://doi.org/10.1002/j.1681-4835.2009.tb00267.x

Carter, L., & Bélanger, F. (2005). The utilization of e-government services: Citizen trust, innovation and acceptance factors. Information Systems Journal, 15(1), 5–25. https://doi.org/10.1111/j.1365-2575.2005.00183.x

Chipeta, J. (2018). A Review of E-government Development in Africa A case of Zambia. Journal of E-Government Studies and Best Practices, 2018, 2155–4137. https://doi.org/10.5171/2018.973845

Danish, D. (2006). The Failure of E-Government in Developing Countries: A Literature Review. The Electronic Journal of Information Systems in Developing Countries, 26(7), 1–10.

Defitri, S. Y., Bahari, A., Handra, H., & Febrianto, R. (2020). Determinant factors of e-government implementation and public accountability: Toe framework approach. Public Policy and Administration, 19(4), 37–51. https://doi.org/10.13165/VPA-20-19-4-03

Dewa, M., & Zlotnikova, I. (2014). C itizens ' Readiness for e-Government Services in Tanzania. Advances in Computer Science: An International Journal, 3(4), 37–45.

Dhillon, R., & Laxmi, V. (2015). Analysis the impact of E-governance Services in Rural Areas of Mansa District-Punjab. International Journal of Computer Science and Information Technologies, 6(5), 4696–4698.

Dhonju, G. R., & Shakya, S. (2019). Analyzing Challenges for the Implementation of E-Government in Municipalities within Kathmandu Valley. Journal of Science and Engineering, 7(November), 70–78. https://doi.org/10.3126/jsce.v7i0.26795

Dias, G. P. (2020). Global e-government development: besides the relative wealth of countries, do policies matter? Transforming Government: People, Process and Policy, 10–18. https://doi.org/10.1108/TG-12-2019-0125

Farzianpour, F., Amirian, S., & Byravan, R. (2015). An Investigation on the Barriers and Facilitators of the Implementation of Electronic Health Records ( EHR ). Scientific Research Publishing, Health(7), 1665–1670. https://doi.org/10.4236/health.2015.712180

Fasheyitan, A. O. (2019). Electronic Government : an Investigation of Factors Facilitating and Impeding the Development of E-Government in Nigeria . August, 402. https://repository.cardiffmet.ac.uk/handle/10369/11156

Haider, Z., Chen, S., Lalani, F., & Mangi, A. A. (2015). Adoption of e-Government in Pakistan: Supply Perspective. International Journal of Advanced Computer Science and Applications, 6(6), 55–63.

Hanum, S., Adawiyah, R. Al, & Sensuse, D. I. (2020). Factors Influencing e-Government Adoption ( A Case Study of Information System Adoption in PPATK ) ( Studi Kasus Adopsi Sistem Informasi di PPATK ). 22(1), 19–30.

Heeks, R. (2003). Most e-government for development projects fail. IGovernment Working Paper Series. http://unpan1.un.org/intradoc/groups/public/documents/NISPAcee/UNPAN015488.pdf

Heeks, R. (2006). Understanding and Measuring eGovernment: International Benchmarking Studies. UNDESA Workshop E-Participation and E-Government: Understanding the Present and Creating the Future, Budapest, July, 27–28.

Herdiyanti, A., Adityaputri, A. N., & Astuti, H. M. (2018). ScienceDirect Understanding the Quality Gap of Information Technology Services from the Perspective of Service Provider and Consumer. Procedia Computer Science, 124, 601–607. https://doi.org/10.1016/j.procs.2017.12.195

Humphrey, A., Paul, K. J., & Mayoka, K. G. (2016). Factors Affecting E-Government Service Utilization in Developing Countries. 5(September), 11–19.

Ibrahim, R., Hilles, S. M. S., Adam, S. M., & El-ebiary, Y. (2016). Methodological Process for Evaluation of E-government Services base on the Federal Republic of Nigerian Citizen ' s E-government Services usage. 9(July). https://doi.org/10.17485/ijst/2016/v9i28/87928

Idoughi, D., & Abdelhakim, D. (2018). Developing countries e-government services evaluation identifying and testing antecedents of satisfaction Case of Algeria. International Journal of Electronic Government Research, 14(1), 63–85. https://doi.org/10.4018/IJEGR.2018010104

Irani, Z. (2014). E-government Implementation Benefits, Risks, and Barriers in Developing Countries: Evidence From Nigeria. 4(1), 13–25.

Jacob, D. W., Fudzee, M. F. M., Salamat, M. A., & Herawan, T. (2019). A review of the generic end-user adoption of e-government services. International Review of Administrative Sciences, 85(4), 799–818. https://doi.org/10.1177/0020852319861895

Janowski, T. (2015). Digital government evolution: From transformation to contextualization. Government Information Quarterly, 32(3), 221–236. https://doi.org/10.1016/j.giq.2015.07.001

Kalu, O. E., & Masri, R. (2019). Challenges of egovernment implementation in the nigerian public service. International Transaction Journal of Engineering , Management , & Applied Sciences & Technologies 10(1), 13–25. https://doi.org/10.14456/ITJEMAST.2019.2

Kanaan, R. K., Abu Hussein, A. M., & Abumatar, G. (2019). Exploring the Factors that affect E-government Implementation in Jordan over Time. Journal of Business & Management (COES&RJ-JBM), 7(3), 252–263. https://doi.org/10.25255/jbm.2019.7.3.252.263

Khadaroo, I., Wong, M. S., & Abdullah, A. (2013). Barriers in local e-government partnership: Evidence from Malaysia. Electronic Government, 10(1), 19–33. https://doi.org/10.1504/EG.2013.051274

Khaemba, S. N., Muketha, G. M., & Matoke, N. (2017). E-Government Systems in Kenya. Journal of Research in Engineering and Applied Sciences, 2(02).

Khamis, M. M., & Weide, T. P. Van Der. (2017). Conceptual Diagram Development for Sustainable e-Government Implementation. 15(1), 33–43.

Khan, F. A., & Ahmad, B. (2015). Factors Influencing Electronic Government Adoption : Perspectives Of Less Frequent Internet Users Of Pakistan. International Journal Of Scientific & Technology Research, 4(01), 306–315.

Kuah, A., Chia, J., & Kuan, L. (2017). Singapore's smart nation initiative – A policy and organisational perspective. Lee Kuan Yew School of Public Policy Paper, 1–12.

Kumar, V., Mukerji, B., Butt, I., & Persaud, A. (2007). Factors for successful e-government adoption: a conceptual framework. Electronic Journal of E-Government, 5(1), 63–76. http://issuu.com/academic-conferences.org/docs/ejeg-volume5-issue1-article89

Layne, K., & Lee, J. (2001). Developing fully functional E- government: A four stage model. Government Information Quarterly, 18(2), 122–136.

Le Blanc, D., & Settecasi, N. (2020). E-participation: a quick overview of recent qualitative trends. United Nations: Department of Economic and Social Affairs Working Paper, 163(163), 1–33. https://www.un.org/development/desa/CONTENTS

Lee, J. B., & Porumbescu, G. A. (2019). Engendering inclusive e-government use through citizen IT training programs. Government Information Quarterly, 36(1), 69–76. https://doi.org/10.1016/j.giq.2018.11.007

Lessa, L. (2019). Sustainability framework for e-government success: Feasibility assessment. ACM International Conference Proceeding Series, Part F1481, 231–239. https://doi.org/10.1145/3326365.3326396

Li, Y., & Shang, H. (2019). Information & Management Service quality , perceived value , and citizens ' continuous-use intention regarding e-government : Empirical evidence from China. Information & Management, July, 103197. https://doi.org/10.1016/j.im.2019.103197

Lindgren, I., Madsen, C. Ø., Hofmann, S., & Melin, U. (2019). Close encounters of the digital kind: A research agenda for the digitalization of public services. Government Information Quarterly, 36(3), 427–436. https://doi.org/10.1016/j.giq.2019.03.002

Lindgren, I., Veenstra, A. F. Van, Lindgren, I., & Veenstra, A. F. Van. (2018). Digital government transformation : a case illustrating public e-service development as part of public sector transformation.

Madariaga, L., Nussbaum, M., Marañón, F., Alarcón, C., & Naranjo, M. A. (2019). User experience of government documents: A framework for informing design decisions. Government Information Quarterly, 36(2), 179–195. https://doi.org/10.1016/j.giq.2018.12.005

Mawela, T., Ochara, N. M., & Twinomurinzi, H. (2017). E-Government Implementation : A Reflection on South African Municipalities. 29(July), 147–171.

Mees, H. L. P., Uittenbroek, C. J., & Driessen, P. P. J. (2019). From citizen participation to government participation : An exploration of the roles of local governments in community initiatives for climate change adaptation in the Netherlands. January, 198–208. https://doi.org/10.1002/eet.1847

Meiyanti, R., Utomo, B., Sensuse, D. I., & Wahyuni, R. (2019). E-Government Challenges in Developing Countries: A Literature Review. 2018 6th International Conference on Cyber and IT Service Management, CITSM 2018, August. https://doi.org/10.1109/CITSM.2018.8674245

Mensah, R., Cater-Steel, A., & Toleman, M. (2020). Factors affecting e-government adoption in Liberia: A practitioner perspective. Electronic Journal of Information Systems in Developing Countries, April, 1–15. https://doi.org/10.1002/isd2.12161

Mercy, M. (2013). E-government project failure in Africa: Lessons for reducing risk. African Journal of Business Management, 7(32), 3196–3201. https://doi.org/10.5897/AJBM12.1093

Mergel, I., Edelmann, N., & Haug, N. (2019). Defining digital transformation : Results from expert interviews ☆. Government Information Quarterly, September 2018, 101385. https://doi.org/10.1016/j.giq.2019.06.002

Mkude, C., Wimmer, M., Mkude, C., Wimmer, M., Framework, S., Developing, E., Mkude, C. G., & Wimmer, M. A. (2017). Strategic Framework for Designing E-Government in Developing

Countries To cite this version : HAL Id : hal-01490902 Strategic Framework for Designing E-Government in Developing Countries.

Mohammed, M., & Hakizimana, W. G. (2019). Investigating challenges in the implementation of e-government services : A case of Rwanda.

Mohlameane, M., & Ruxwana, N. (2020). Exploring the impact of cloud computing on existing South African regulatory frameworks. SA Journal of Information Management, 22(1), 1–9. https://doi.org/10.4102/sajim.v22i1.1132

Muhammad, M. (2013). Managing the implementation of e-government in Malaysia: a case of E-Syariah. Australian Journal of Basic and Applied Sciences, 7(8), 92–99.

Mukamurenzi, S. (2019). Improving qualities of e - government services in Rwanda : A service provider perspective. February, 1–16. https://doi.org/10.1002/isd2.12089

Munyoka, W. (2019). Electronic government adoption in voluntary environments – a case study of Zimbabwe. July. https://doi.org/10.1177/0266666919864713

Nabafu, R., & Maiga, G. (2012). A Model of Success Factors for Implementing Local E-government in Uganda. e-Government in Sub Sahara African Countries Local e-Government. 10(1), 31–46.

Nakakawa, A., & Namagembe, F. (2019). Requirements for developing interoperable e-government systems in developing countries - A case of Uganda. Electronic Government, 15(1), 67–90. https://doi.org/10.1504/EG.2019.096577

Napitupulu, D., Syafrullah, G. M., Rahim, R., Ahmar, A. S., & Sucahyo, Y. (2018). Content validity of critical success factors for e-Government implementation in Indonesia. IOP Conference Series: Materials Science and Engineering, 352(1), 012058. https://doi.org/10.1088/1757-899X/352/1/012058

Nawafleh, S., Obiedat, R., & Harfoushi, O. (2012). E-Government Between Developed and Developing Countries. International Journal of Advanced Corporate Learning (IJAC), 5(1). https://doi.org/10.3991/ijac.v5i1.1887

Nhema, A. G. (2016). E-Government and Development in Zimbabwe : An Appraisal. 6(2), 13–23.

Niehaves, B., & Becker, J. (2008). The age-divide in E-government - Data, interpretations, theory fragments. IFIP International Federation for Information Processing, 286, 279–287. https://doi.org/10.1007/978-0-387-85691-9_24

Nurdin, N., Stockdale, R., & Scheepers, H. (2011). Understanding organizational barriers influencing local electronic government adoption and implementation: The electronic government implementation framework. Journal of Theoretical and Applied Electronic Commerce Research, 6(3), 13–27. https://doi.org/10.4067/S0718-18762011000300003

Nurdin, N., Stockdale, R., & Scheepers, H. (2014). Coordination and cooperation in e-government: An Indonesian local e-government case. Electronic Journal of Information Systems in Developing Countries, 61(1), 1–21. https://doi.org/10.1002/j.1681-4835.2014.tb00432.x

Ohemeng, F. L. K., & Ofosu-Adarkwa, K. (2014). Overcoming the Digital Divide in Developing Countries: An Examination of Ghana's Strategies to Promote Universal Access to Information Communication Technologies (ICTs). Journal of Developing Societies, 30(3), 297–322. https://doi.org/10.1177/0169796X14536970

Ojha, S., & Pandey, I. M. (2017). Management and financing of e-Government projects in India: Does financing strategy add value? IIMB Management Review, 29(2), 90–108. https://doi.org/10.1016/j.iimb.2017.04.002

Owusu-Ansah, S. (2014). Human Factor Issues In The Use Of E-Government Services Among Ghanaian Middle Age Population : Improving Usability Of Existing And Future Government Virtual Interactive Systems Design. Journal of Information Engineering and Applications, 4(4), 85–107.

Patsioura, F. (2014). Evaluating e-government. Evaluating Websites and Web Services: Interdisciplinary Perspectives on User Satisfaction, 1–18. https://doi.org/10.4018/978-1-4666-5129-6.ch001

Pederson, K. (2016). e-Government in Local Government: Challenges and Capabilities. Electronic Journal of E-Government, 14(1), 99–116. https://search.proquest.com/docview/1804472280?accountid=17242

Pérez-Morote, R., Pontones-Rosa, C., & Núñez-Chicharro, M. (2020). The effects of e-government evaluation, trust and the digital divide in the levels of e-government use in European countries. Technological Forecasting and Social Change, 154(January), 119973. https://doi.org/10.1016/j.techfore.2020.119973

Rabaa, A. A., Zogheib, B., Alshatti, A., & Jamal, E. Al. (2018). Adoption of e-Government in Developing Countries : The Case of the State of Adoption of e-Government in Developing Countries : The Case of the State of Kuwait. June.

Ramdan, S. M. ., Azizan, Y. N., & Saadan, K. (2014). E-Government Systems Success Evaluating under Principle Islam: A Validation of the Delone and Mclean Model of Islamic Information Systems Success. Academic Research International, 5(2), 72–85. www.savap.org.pk%0Awww.journals.savap.org.pk

Regmi, N. (2017). Expectations versus reality: A case of internet in Nepal. Electronic Journal of Information Systems in Developing Countries, 82(1), 1–20. https://doi.org/10.1002/j.1681-4835.2017.tb00607.x

Richardson, J. W. (2011). Challenges of adopting the use of technology in less developed countries: The case of Cambodia. Comparative Education Review, 55(1), 8–29.

Rowley, J. (2011). E-Government stakeholders - Who are they and what do they want? International Journal of Information Management, 31(1), 53–62. https://doi.org/10.1016/j.ijinfomgt.2010.05.005

Sánchez-Torres, J. M., & Miles, I. (2017). The role of future-oriented technology analysis in e-Government: a systematic review. European Journal of Futures Research, 5(1). https://doi.org/10.1007/s40309-017-0131-7

Sareen, M., Punia, D. K., & Chanana, L. (2013). Exploring factors affecting use of mobile government services in India. Problems and Perspectives in Management, 11(4), 86–93.

Sarrayrih, M. A., & Sriram, B. (2015). Major challenges in developing a successful e-government: A review on the Sultanate of Oman. Journal of King Saud University - Computer and Information Sciences, 27(2), 230–235. https://doi.org/10.1016/j.jksuci.2014.04.004

Sharma, G., Bao, X., & Peng, L. (2014). Public Participation and Ethical Issues on E-governance: A Study Perspective in Nepal. Ejeg.Com, 12(1), 82–96. http://www.ejeg.com/issue/download.html?idArticle=312

Sikaonga, S., & Tembo, S. (2020). E-Government Readiness in the Civil Service : A Case of Zambian Ministries. International Journal of Information Science 2020, 10(1), 15–28. https://doi.org/10.5923/j.ijis.20201001.03

Singh, S., & Travica, B. (2018). E-Government systems in South Africa: An infoculture perspective. Electronic Journal of Information Systems in Developing Countries, July, 1–16. https://doi.org/10.1002/isd2.12030

Singh, Shawren, & Travica, B. (2018). E-Government systems in South Africa : An infoculture perspective. July, 1–16. https://doi.org/10.1002/isd2.12030

Sterrenberg, G., & Keating, B. (2016). Measuring IS success of e-government : A Case Study on the Disability Sector in Australia. Australasian Conference on Information Systems, 2003, 1–8.

Sulehat, N. A., & Taib, D. C. A. (2016). E-Government Information Systems Interoperability in Developing Countries. Journal of Business and Social Review in Emerging Economies, 2(1), 49–60. https://doi.org/10.26710/jbsee.v2i1.18

Susanto, T. D., & Aljoza, M. (2015). Individual Acceptance of e-Government Services in a Developing Country: Dimensions of Perceived Usefulness and Perceived Ease of Use and the Importance of Trust and Social Influence. Procedia Computer Science, 72, 622–629. https://doi.org/10.1016/j.procs.2015.12.171

Tan, B., & Zhou, Y. (2018). Technology and the City: Foundation for a Smart Nation. https://www.clc.gov.sg/research-publications/framework

Tirastittam, P., Sotarat, T., & Chuckpaiwong, R. (2018). A Study of Bureaucracy in the Digital Transformation Era: A Global Organizational Context. ITMSOC Transactions on Innovation & Business Engineering, 03, 30–34.

Twizeyimana, J. D., & Andersson, A. (2019a). The public value of E-Government – A literature review. Government Information Quarterly, 36(2), 167–178. https://doi.org/10.1016/j.giq.2019.01.001

Twizeyimana, J. D., & Andersson, A. (2019b). The public value of E-Government – A literature review. Government Information Quarterly, 36(2), 167–178. https://doi.org/10.1016/j.giq.2019.01.001

Twizeyimana, J. D., Larsson, H., & Grönlund, Å. (2018). E-government in Rwanda: Implementation, Challenges and Reflections. 16(1), 19–31.

Van Deursen, A., & Van Dijk, J. (2010). Civil servants' internet skills: Are they ready for e-government? Lecture Notes in Computer Science (Including Subseries Lecture Notes in Artificial Intelligence and Lecture Notes in Bioinformatics), 6228 LNCS, 132–143. https://doi.org/10.1007/978-3-642-14799-9_12

Vejačka, M. (2018). Acceptance of e-government services by business users: The case of Slovakia. Journal of Applied Economic Sciences.

Venkatesh, V., Sykes, T. A., & Venkatraman, S. (2014). Understanding e-Government portal use in rural India: Role of demographic and personality characteristics. Information Systems Journal, 24(3), 249–269. https://doi.org/10.1111/isj.12008

Verkijika, S. F. (2018). Quality assessment of e - government websites in Sub - Saharan Africa : A public values perspective. 1–17. https://doi.org/10.1002/isd2.12015

Verkijika, S. F., & De Wet, L. (2018). Quality assessment of e-government websites in Sub-Saharan Africa: A public values perspective. Electronic Journal of Information Systems in Developing Countries, 84(2), 1–17. https://doi.org/10.1002/isd2.12015

Voutinioti, A. (2014). Determinants of user adoption of e-government services: the case of Greek Local Government. International Journal of Technology Marketing, 9(3), 234. https://doi.org/10.1504/ijtmkt.2014.06385

Vrabie, C. (2012). Computer Competencies Necessary for an Effective E-Governance. Governance, July 2015.

Weller, S. C., Vickers, B., Bernard, H. R., Blackburn, A. M., Borgatti, S., Gravlee, C. C., & Johnson, J. C. (2018). Open-ended interview questions and saturation. 1–18.

Yang, Y. (2017). Towards a new digital era: Observing local E-Government services adoption in a Chinese municipality. Future Internet, 9(3). https://doi.org/10.3390/fi9030053

Yin, R. K. (2009). Case Study Research. Design and Methods. Sage, 4(4), 264–267. https://doi.org/10.1007/BF01103312

Zaied, A., Khairalla, F., & Al-Rashed, W. (2007). Assessing e-readiness in the Arab countries: Perceptions towards ICT environment in public organisations in the State of Kuwait. The Electronic Journal of E-Government, 5(1), 77–86.

Zaied, A. N. H., Ali, A. H., & El-ghareeb, H. A. (2017). E-government Adoption in Egypt : Analysis, Challenges and Prospects. 52(2), 70–79.

Zautashvili, D. (2018). E-government Maturity Model by Growth Level of E-services Delivery. Journal of Technical Science and Technologies, 6(2), 17–22.

Ziba, P. W., & Kang, J. (2020). Factors affecting the intention to adopt e-government services in Malawi and the role played by donors. Information Development, 36(3), 369–389. https://doi.org/10.1177/0266666919855427